# First-principles investigation of boron defects in nickel ferrite spinel

Zs. Rák[1], C. J. O'Brien, and D. W. Brenner

Department of Materials Science and Engineering, North Carolina State University,

Raleigh, 27695-7907

**Abstract**

The accumulation of boron within the porous nickel ferrite (NiFe$_2$O$_4$, NFO) deposited on fuel rods is a major technological problem with important safety and economical implications. In this work the electronic structure of nickel ferrite is investigated using first-principles methods, and the results are combined with experimental data to analyze B incorporation into the NFO structure. Under thermodynamic equilibrium the calculations predict that the incorporation of B into the NFO structure is unfavorable. The main limiting factors are the narrow stability domain of NFO and the precipitation of B$_2$O$_3$, Fe$_3$BO$_5$, and Ni$_3$B$_2$O$_6$ as secondary phases. In n-type NFO, the most stable defect is Ni vacancy $\left(V_{Ni}^{2-}\right)$ while in p-type material lowest the formation energy belongs to tetrahedrally coordinated interstitial B $\left(B_T^{2+}\right)$. Because of these limiting conditions it is more thermodynamically favorable for B to form secondary phases with Fe, Ni and O than it is to form point defects in NFO.

## 1. Introduction

Because of the increasing cost of fossil fuel based energy and the pressure to reduce greenhouse gases, energy policy in the United States currently encourages research and development in nuclear energy [1]. To allow longer lifetimes and higher power output from a nuclear reactor, a major technical issue that has to be conquered is the corrosion and corrosion-related failure of nuclear fuel. CRUD (Chalk River Unidentified Deposits) is the name given to corrosion products that accumulate on the hot surface of the fuel cladding. In a pressurized water reactor (PWR) CRUD is produced from dissolved metal cations and particulate corrosion products originating from the surfaces of the reactor coolant system, and consists mainly of nickel ferrite spinel ($NiFe_2O_4$, NFO), nickel oxide (NiO).[2, 3] Although a thin layer of CRUD on the cladding surface can enhance heat transfer, thicker deposits have a negative effect on fuel performance and generate operational challenges. Thick CRUD, usually deposited on the upper part of the fuel rods, reduces the heat transfer in the reactor core and raises the local surface temperature of the cladding, which increases the corrosion rate. Furthermore, boron (B) can accumulate in thick CRUD, which can trigger fluctuations in the neutron flux and cause a shift in the power output from the top half to the bottom half of the core. This phenomenon, known as the axial offset anomaly (AOA), has important safety implications and could lead to down-rating of a power plant with significant economic consequences. Therefore, it is crucial to understand and predict the mechanisms of CRUD formation and B accumulation within the CRUD.

[1] Corresponding author:
Department of Materials Science and Engineering,
North Carolina State University,
911 Partners Way, EB 1 BOX 7907
Raleigh, NC 27695-7907
Email: zrak@ncsu.edu (Zs. Rak)

Even though it is clear that the key factors necessary for AOA to manifest are (i) sub-cooled nucleate boiling, (ii) thick CRUD, and (iii) sufficient B in the coolant, many details about the root cause of AOA are not fully understood. For example, little is known about the role played by the coolant chemistry in AOA development or about the precise mechanism of B deposition. Within the EPRI/WEC (Electric Power Research Institute/Westinghouse Electric Company) simulation models B uptake by CRUD is considered to take place *via* lithium borates in the form of $LiBO_2$, $Li_2BO_7$, and $Li_2B_4O_7$.[4-7] However, because these compounds exhibit retrograde solubility with respect to temperature, they are difficult to observe in PWR CRUD scrapes, but they have been observed in simulated CRUD.[8] Sawicki analyzed the corrosion deposits found on the fuel assemblies in various PWRs and identified the formation of $Ni_2FeBO_5$ (mineral bonnacordite) as a new mechanism for B retention.[9, 10] The results of Sawicki's work have triggered new modeling efforts to describe the chemistry of borated fuel CRUD.[11] Mesoscale CRUD models under development within the CASL program (Consortium for Advanced Simulation of Light Water Reactors) assume precipitation of boron oxide ($B_2O_3$) as a possible B deposition mechanism in the porous CRUD.[12]

Although current research is moving toward a better description of CRUD properties and its behavior in PWRs, more work is needed to elucidate the mechanisms through which B or boron-containing solids are deposited on the fuel rods. In this work we carry out electronic structure calculations to investigate the possibility of B incorporation into crystalline NFO. To do this, we treat B as either a substitutional or interstitial point defect in the NFO structure and use first-principles-based thermodynamics to analyze the stability of B inside the structure.

## 2. Crystal and defect modeling

The spinel atomic arrangement is shared by many transition metal oxides with formula $AB_2O_4$ (*Fd3m*), where A and B are divalent ($A^{2+}$) and trivalent ($B^{3+}$) ions, respectively. In the normal structure the oxygen ions form a face centered cubic array and the $A^{2+}$ and $B^{3+}$ ions sit in tetrahedral (1/8 occupied) and octahedral (1/2 occupied) sites in the lattice, giving a unit cell with 8 A's, 16 B's and 32 O's. The inverse spinel is an alternative arrangement where the divalent ions swap with half of the trivalent ions so that the $A^{2+}$ ions occupy octahedral sites. Therefore, the general formula of an inverse spinel can be written as $B^{3+}(A^{2+}B^{3+})O_4$.

Nickel ferrite crystallizes in the inverse spinel structure, with half of the $Fe^{3+}$ ions occupying the tetrahedral sites while $Ni^{2+}$ and the remaining $Fe^{3+}$ are randomly distributed over the octahedral sites. Because of the periodic boundary conditions employed in our calculations, any atomic distribution within a supercell corresponds to an array of cations with long-range order. In theory, the cation distribution could be modeled using a special quasi-random structure;[13, 14] however, the large unit cells required for such simulation would lead to prohibitively expensive computational efforts. In this work we employ the structural model used by Fritsch and Ederer, where the Ni and Fe cations are distributed over the octahedral sites such that the symmetry is reduced from cubic (*Fd3m*) to orthorhombic (*Imma*).[15] The same structure has been used by O'Brien *et al.*[16] to investigate the thermodynamic properties of NFO surfaces under conditions typical to PWR coolant.

The point defects that are investigated in the present work are Ni and Fe vacancies as well as substitutional and interstitial B impurities. The vacancies can be of three types depending on the site from which the atom is removed: tetrahedral and octahedral Fe vacancies ($V_{Fe}^T$ and $V_{Fe}^{O}$), and octahedral Ni vacancy ($V_{Ni}^{O}$). Similarly, depending on which atom is replaced by B, there are three types of substitutional impurities: boron can substitute for a Fe atom at a tetrahedral or an octahedral site ($B_{Fe}^{T}$ and $B_{Fe}^{O}$) or one Ni atom at an octahedral ($B_{Ni}^{O}$). The interstitial B can be located either at an empty octahedral or at an empty tetrahedral site. Assuming a random Ni distribution, all unoccupied octahedral sites are equivalent, while there are two types of unoccupied tetrahedral sites. The three possible interstitial atomic configurations are illustrated in Fig. 1, where the two tetrahedral sites are denoted T1 and T2.

### 3. Computational parameters

The calculations were performed using the projector augmented wave (PAW)[17, 18] method within density functional theory (DFT)[19, 20] as implemented in the Vienna Ab initio Simulation Package (VASP).[21-24] The exchange-correlation potential was approximated by the generalized gradient approximation (GGA), as parameterized by Perdew, Burke, and Ernzerhof (PBE).[25]. The standard PAW potentials, supplied with the VASP package, were employed in the calculations.[17, 18] The 3$d$ and 4$s$ states of Ni and Fe as well as the 2$s$ and 2$p$ states of B and O are considered as valence states while the rest are treated as core states. The cut-off energy for the plane wave basis was set to 550 eV and the convergence of self-consistent cycles was assumed when the energy difference between two consecutive cycle was less than $10^{-4}$ eV. The Brillouin-zone was sampled by the Γ-point in all calculations and a Gaussian smearing of 0.1 eV was used. The internal structural parameters were relaxed until the total energy and the Hellmann-Feynman forces on each nucleus were less than 0.02 eV/Å. To minimize the interaction between periodic images of defects, all defect calculations have been performed on 2×2×2 supercells containing 448 atoms, using the calculated lattice constant of pure NFO (a = 8.41 Å).[26]

To describe the behavior of the localized Fe 3$d$ and Ni 3$d$ states, the orbital-dependent, Coulomb potential (Hubbard $U$) and the exchange parameter $J$ were included in the calculations using the DFT+$U$ formalism.[27] The simplified, rotationally invariant approach introduced by Dudarev *et al*[28] was used. The value of the Hubbard $U$ parameter can be estimated from band-structure calculations in the supercell approximation with different $d$ and $f$ occupations[29] or from calculations based on a constrained random-phase approximation.[27, 30] Here $U$ and $J$ are treated as parameters with values $U(Fe_d)$ = 5.5 eV with the corresponding $J(Fe_d)$ = 1.0 eV, and $U(Ni_d)$ = 7.0 eV with $J(Ni_d)$ = 1.0 eV. These values are physically reasonable and are within the range of the previous values in the literature.[16, 31-33]

### 4. Thermodynamics of B incorporation intro nickel ferrite

The formation of a defect in a crystalline solid can be regarded as an exchange of atoms and electrons between the host material and chemical reservoirs. Therefore the formation energy of a defect $D$ in charge state $q$ can be written as:[34, 35]

$$\Delta H_f(D^q) = \Delta E(D^q) + \sum_i n_i \mu_i + qE_F, \qquad (1)$$

where

$$\Delta E(D^q) = E(D^q) - E_0 + \sum_i n_i E_i + qE_{VBM} \qquad (2)$$

In Eqs. (1) and (2) $E(D^q)$ and $E_0$ are the total energies of the defect-containing and defect free solids. The second term on the right side of Eq. (1) represents the change in energy due to the exchange of particles between the host compound and the chemical reservoir, where $\mu_i$ are the chemical potential of the atomic species $i$ ($i$ = Ni, Fe, or B) referenced to the elemental solid/gas with energy $E_i$ and $n_i$ are the number of atoms added to $(n_i < 0)$ or removed from $(n_i > 0)$ the supercell. The quantity $E_F$ is the Fermi energy referenced to the energy of the valence band maximum (VBM), $E_{VBM}$. This value is calculated as the VBM energy of the pure NFO, corrected by aligning the core potential of atoms far away from the defect in the defect-containing supercell with that in the defect free supercell.[35] The quantity $q$ represents the charge state of the defect, i. e. the number of electrons exchanged with the electron reservoir with chemical potential $E_F$.

Equation (1) shows that, in principle, by adjusting the atomic chemical potential of the constituents and by tuning the electronic Fermi energy, one can control the defect formation energy and, consequently, the solubility of the dopant in the host matrix. Under thermodynamic equilibrium, the achievable values of the chemical potentials are limited by several conditions:

(i) to avoid elemental precipitations, the chemical potentials are bound by

$$\mu_{Ni} \leq 0, \mu_{Fe} \leq 0, \mu_O \leq 0, \text{ and } \mu_B \leq 0 \qquad (3)$$

(ii) to maintain a stable NFO host the $\mu_i$'s must satisfy

$$\mu_{Ni} + 2\mu_{Fe} + 4\mu_O = \Delta H(\text{NiFe}_2\text{O}_4) = \Delta H(\text{NFO}) \qquad (4)$$

where $\Delta H(\text{NFO})$ is the formation enthalpy of $\text{NiFe}_2\text{O}_4$,

(iii) to avoid formation of competing phases, such as iron oxides (Wüstite, hematite and magnetite) and nickel oxides, the following conditions must apply

$$n\mu_{Fe} + m\mu_O \leq \Delta H(\text{Fe}_n\text{O}_m), \text{ where } (n,m) = (1,1), (2,3), \text{ and } (3,4) \qquad (5)$$

$$n\mu_{Ni} + m\mu_O \leq \Delta H(\text{Ni}_n\text{O}_m), \text{ where } (n,m) = (1,1) \text{ and } (2,3) \qquad (6)$$

Further constraints on the chemical potential are posed by the possibility of forming secondary phases between boron and the host elements. In this work we consider the following compounds as possible secondary phases: $B_2O_3$, $NiB$, $Ni_2B$, $Ni_4B_3$, $FeBO_3$, $Fe_3BO_5$, $Fe_3BO_6$, $NiB_2O_4$, $Ni_3B_2O_6$, and $Ni_2FeBO_5$. To avoid the formation of these compounds, the chemical potentials of B, Ni, Fe, and O must satisfy conditions similar to (5) and (6).

## 5. Formation enthalpies and elemental reference energies

The theoretical enthalpy (heat) of formation of a compound $A_nB_m...$ can be calculated as:

$$\Delta H^{theor}(A_nB_m...) = E(A_nB_m...) - nE_A - mE_B - ... \qquad (7)$$

where $E(A_nB_m...)$ is the total energy per formula unit (f. u.) of the compound and $E_A, E_B,...$ are the total energies per atom of the elements in their standard state. According to Eq. (7), to predict the enthalpies of formation required by conditions (ii)-(iii), it is necessary to compute the differences between the total energies of the compound $A_nB_m...$ and their elemental constituents A, B, ... in the standard state. If the compound and the elemental materials are chemically similar, Eq. (7) yields very accurate results because the DFT errors will cancel out when calculating the total energy differences. In our case the compounds are the insulating or semiconducting materials, such as NFO, $Fe_nO_m$, or $Ni_nO_m$, while the elemental phases include the metallic form of the cations (Ni, Fe, and semimetallic B) and the gaseous anion ($O_2$). These are chemically and physically dissimilar systems, where the cancellation of the DFT errors is known to be incomplete.[36, 37] Furthermore, to reproduce the correct electronic structures of the Ni- and Fe-containing systems it is essential to include the Hubbard $U$ correction for the Ni and Fe $3d$ states. The values of the $U$ parameters used for the non-metal compounds, however, are different from those used for metallic phases, which can lead to large errors in the formation enthalpy calculations. One way to overcome these issues is to utilize the experimental enthalpies of formation $\left(\Delta H^{\exp}\right)$ in Eqs. (4)-(6). However, this approach cannot be used because the experimental value of the formation enthalpy of the ternary compounds from the elemental constituents is not available.

To compute the formation enthalpies we use an approach,[16, 32, 38-40] in which the elemental energies $E_A, E_B, ...$ are approximated from the system of linear equations:

$$\Delta H^{\exp}(A_n B_m ...) = E(A_n B_m ...) - nE_A - mE_B - ... \quad (8)$$

We calculate the DFT energies of 10 binary compounds that can be formed from Ni, Fe, O, and B, for which the $\Delta H^{\exp}$ values are available,[41, 42] and then we solve the overdetermined system of equations Eq. (8) in the least-squares approach. This way we obtain the elemental energies, $E_A, E_B, ...$, without directly calculating the DFT energies of the elements in their standard metallic or gaseous state. The obtained values are used as the elemental reference energies to calculate the defect formation energies in Eq. (1) and the enthalpies of formation required in Eqs. (4)-(6). The experimental and theoretical values of the formation enthalpies are listed in Table 1, along with the DFT total energies of the compounds and the fitted elemental reference energies are listed in Table 2.

### 6. Results and discussions

To assess the possibility of B incorporation into NFO, we calculate the electronic structure of NFO, evaluate the formation energies of B-related defects and use the thermodynamic scheme described above to estimate the defect stability with reference to the formation of Ni-Fe-B compounds as secondary phases.

*6.1 Pure NFO*

Nickel ferrite spinel is a ferrimagnetic insulator with high Curie temperature (870K)[43] and, therefore, it has a great potential for technological applications especially in the area of spintronics.[44] However, because in this work we do not focus on device applications of NFO, we only give a brief description of the calculated electronic structure of NFO, comparing our results with data available in the literature.

The total electronic density of states (DOS), along with the DOS projected on the *d*-states of Fe and Ni ions are illustrated in Fig. 2. Because the Fe ions in NFO are in the 3+ oxidation state, the Fe 3*d* shells are half filled. Fig. 2 (a) and (b) reveal that all Fe's are in the high spin state, with one spin projection completely occupied (states between -6.5 and -8.0 eV) and the other spin projection completely empty (states between 1.5 and 3.0 eV). The Fe *d* states are strongly localized; they are separated from other valence band (VB) and conduction band (CB) states. In contrast, the strong hybridization between the Ni 3*d* and O 2*p* states (Fig. 2 (c)), produces a VB that displays both Ni *d* and O *p* character across the energy range from -6.0 eV to $E_F$. The hybridization takes place mainly between the Ni $t_{2g}$ and O 2*p* states, while the Ni $e_g$ states remain more separated and localized.

Figure 2 also illustrates the origin of the ferrimagnetism in NFO; the magnetic moments of the octahedral $Fe^{3+}$ and $Ni^{2+}$ cations are parallel and they couple antiferromagnetically with the tetrahedral $Fe^{3+}$ moments. Given that there are equal numbers of $Fe^{3+}$ on the octahedral and tetrahedral sites, the $Fe^{3+}$ magnetic moments are compensated, so the net moment is attributable mainly to the octahedral $Ni^{2+}$ cations. The calculated spin magnetic moments, listed in Table III, are consistent with this magnetic structure and are in fairly good agreement with earlier experimental[45] and theoretical values.[15, 46]

The calculated band gap measured from the VBM to the conduction band minimum (CBM) is 1.3 eV, somewhat larger than the gaps of 0.97, 0.99. and 1.1 eV, obtained with the DFT+*U* method by Fritsch and Ederer[15], Antonov *et al.*[46], and Sun *et al.*[47], respectively. The slightly increased band gap in our calculations is the result of a larger on-site Coulomb interaction ($U_{eff}$ parameter) applied to the Ni *d* states compared with previously used values in the literature.[15, 46, 47] Because the VBM displays a mixture of O *p* and Ni *d* character, an increase in the Ni *d-d* correlation shifts the occupied Ni *d* states, and therefore the VBM, to lower energies relative to the CBM.

*6.2 Stability of NFO*

As described in Section 4, the achievable values of Ni, Fe and O chemical potentials are bound by the conditions to maintain a stable NFO and avoid formation of competing phases (including elemental solids/gases). The calculated chemical potential domain based on Eqs. (3)-(6), where NFO is stable, is illustrated by the dark area in Fig. 3. Under thermodynamic equilibrium, the vertices of the triangle in Fig. 3 represent the achievable limits of the Ni and Fe chemical potentials; A corresponds to the Fe-rich, Ni-rich ($\mu_{Fe} = \mu_{Ni} = 0$) limit, B corresponds to the Ni-rich, Fe-poor ($\mu_{Ni} = 0, \mu_{Fe} = -5.71\,\text{eV}$) limit, and C is the Ni-poor, Fe-rich ($\mu_{Ni} = -11.411\,\text{eV}, \mu_{Fe} = 0$) limit. As is apparent from Fig. 3, the domain of the allowed chemical potentials for stable NFO is relatively narrow. In the white areas, NFO is unstable with respect to competing phases; the lower part of the triangle represents the region where iron oxides (FeO, $Fe_2O_3$, and $Fe_3O_4$) form, while the upper region is excluded due to precipitation of nickel oxides (NiO and $Ni_2O_3$). Figure 3 also illustrates that under Fe- or Ni-rich conditions ($\mu_{Fe} = 0$ or $\mu_{Fe} = 0$) NFO is not stable. The highest achievable values of $\mu_{Fe}$ and $\mu_{Ni}$ for stable NFO are defined by the intersection of the lines that set the limit for $Fe_3O_4$ and NiO in the ($\mu_{Fe}, \mu_{Ni}$) plane. At this point, represented as X in Fig. 3, the calculated Fe and Ni chemical potentials are $\mu_{Fe} = -1.09\,\text{eV}$ and $\mu_{Ni} = -0.70\,\text{eV}$. According to Eq. (4), at point X, the O chemical potential is $\mu_O = -2.13\,\text{eV}$. This is also the lowest possible value of $\mu_O$ that that still assures a stable NFO.

The lowest achievable values of $\mu_{Fe}$ and $\mu_{Ni}$ for stable NFO are defined by the intersection of the O-rich line (BC segment on Fig. 3) with the lines that limit the formation NiO and $Fe_2O_3$, respectively. These intersections are denoted as Y and Z on Fig. 3. The calculated values of the chemical potentials at Y and Z are $\mu_{Fe} = -4.29\,\text{eV}$, $\mu_{Ni} = -2.83\,\text{eV}$ and $\mu_{Fe} = -4.20\,\text{eV}$, $\mu_{Ni} = -3.01\,\text{eV}$, respectively.

*6.3 Fe and Ni vacancies*

According to Eq. (1) and (2) the formation energy of a defect $D$ in charge state $q$ is equal to the difference between the energies of the defect-containing and defect free supercells, corrected by the chemical potentials of the atomic and electronic reservoirs with which the system exchanges particles (atoms and electrons). In the case of Fe or Ni vacancies ($V_{Fe/Ni}$) the formation energies are given by $\Delta H_f\left(V_{Fe/Ni}^q\right) = \Delta E\left(V_{Fe/Ni}^q\right) + \mu_{Fe/Ni} + qE_F$. The first term on the right-hand side represents the energy difference between the vacancy-containing and pristine systems, corrected by the elemental reference energies and VBM energy (Eq. (2)). The calculated values of $\Delta E\left(D^q\right)$, listed in Table 4, can be used to assess the defect formation in NFO under the conditions defined by (i)-(iii).

To make the formation of Fe/Ni vacancy favorable, we have to create Fe/Ni-poor conditions, that is we have to minimize the chemical potential of Fe/Ni. As calculated in the previous section, the lowest possible values of $\mu_{Fe}$ and $\mu_{Ni}$ that satisfies conditions (i) - (iii) are -4.29 and -3.01 eV, respectively. At these chemical potentials the formation energies of tetrahedral and octahedral Fe and octahedral Ni vacancies in neutral charge states are $\Delta H_f\left(V_{Fe^T}^0\right)$ = 7.73 – 4.29 = 3.44 eV, $\Delta H_f\left(V_{Fe^O}^0\right)$ = 7.46 – 4.29 = 3.17 eV, and $\Delta H_f\left(V_{Ni^O}^0\right)$ = 5.12 – 3.01 = 2.11 eV. These values are moderately high, indicating that formation of neutral vacancies in NFO is unlikely.

In the case of the charged defects, the formation energy depends on the Fermi level, because to ionize a defect, electrons must be added to or taken from an electron reservoir with energy $E_F$. Figure 4 illustrates the vacancy formation energies as a function of $E_F$, calculated for different charge states, under Fe-poor and Ni-poor conditions. We observe that the Ni and Fe vacancies display similar behavior; when $E_F$ is tuned closer to the VBM or CBM, the vacancies become charged and the formation energies decrease considerably. This effect is stronger in n-type NFO ($E_F$ closer to CBM), where the formation energy of Ni vacancy drops to approximately 0.5 eV. Figure 4 also illustrates that for all values of $E_F$ within the band gap, the formation energy of Ni vacancy is the lowest, suggesting that under appropriate conditions, Ni vacancy could be the dominant intrinsic defect in NFO.

*6.4 Substitutional B defects*

There are three types of substitutional B defects in NFO; boron can occupy a tetrahedral or octahedral Fe site ($B_{Fe}^T$ or $B_{Fe}^O$) or it can substitute for an octahedral Ni ion ($B_{Ni}^O$). Given that the methodology of the formation energy calculations are the same for the Fe- and Ni-site defects, here we describe the details of Fe-site defect calculations while for the Ni-site defect we only present the results.

The formation energy of substitutional B at the Fe site is given by $\Delta H_f\left(B_{Fe^{T/O}}^q\right) = \Delta E\left(V_{Fe^{T/O}}^q\right) - \mu_B + \mu_{Fe} + qE_F$. Even though the O and Ni chemical potentials do not appear explicitly in this expression, the formation energy depends indirectly on $\mu_{Ni}$ and $\mu_O$ through conditions (ii) and (iii), as described in Section 4. To make B incorporation energetically favorable, $\mu_{Fe}$ has to be minimized and $\mu_B$ maximized As described earlier, the lowest possible value of $\mu_{Fe}$ that maintains a stable NFO is -4.29 eV, represented by the Y point on Fig. 3. At this point the chemical potentials of O and Ni are zero and -2.83 eV, respectively. Using these values in setting up the conditions to avoid formation of secondary phases, we find that the highest possible value of $\mu_B$ is -6.60 eV due to the restriction to avoid $Ni_3B_2O_6$. Under these conditions, the formation energies of neutral $B_{Fe}^T$ and $B_{Fe}^O$ are $\Delta H_f\left(B_{Fe^T}^0\right) = -0.08 + 6.60 - 4.29 = 2.23$ eV and $\Delta H_f\left(B_{Fe^O}^0\right) = 1.93 + 6.60 - 4.29 = 4.24$ eV. These values are relatively high, indicating that, under Fe-poor conditions, B incorporation at the Fe site in NFO is limited by the formation of $Ni_3B_2O_6$ as a secondary phase.

Because the formation energies of substitutional B depend indirectly on $\mu_{Ni}$ and $\mu_O$, instead of limiting the investigation to Fe-poor conditions, we have to explore the entire range of chemical potentials where NFO is stable. Within the limits of condition (iii), reduced values of $\mu_{Ni}$ and $\mu_O$ allow for larger $\mu_B$ which, in turn, decreases the formation energies. The minimal value of $\mu_O$ (O-poor condition) is obtained at the X point in Fig. 3 and it is -2.13 eV. At this point $\mu_{Fe}$ = -1.09 eV and $\mu_{Ni}$ = -0.70 eV. The maximum value of $\mu_B$ is obtained by imposing condition (iii) on all possible secondary phases. In this case, the limiting condition is given by $Fe_3BO_5$ resulting in a value of $\mu_B$ that must be less than -3.63 eV. Therefore, under O-poor conditions the formation energies of neutral $B_{Fe}^T$ and $B_{Fe}^O$ are $\Delta H_f\left(B_{Fe^T}^0\right) = -0.08 + 3.63 - 1.09 = 2.46$ eV and $\Delta H_f\left(B_{Fe^O}^0\right) = 1.93 + 3.63 - 1.09 = 4.47$ eV. Similar calculations can be carried out under Ni-poor conditions, corresponding to point Z in Fig. 3, where $\mu_{Ni}$ = -3.01 eV, $\mu_{Fe}$ = -4.20 eV, and $\mu_O$ = 0. In this case the chemical potential of B is $\mu_B \leq$ -6.42 eV, and limited by the formation of $B_2O_3$. The formation energies of substitutional B impurities at Ni-poor conditions are calculated as $\Delta H_f\left(B_{Fe^T}^0\right) = 2.14$ eV and $\Delta H_f\left(B_{Fe^O}^0\right) = 4.15$ eV.

In the case of the substitutional B impurity at the Ni site, after exploring the entire stability domain of NFO, the lowest defect formation energy was obtained at Fe-rich conditions (point X in Fig. 3), where $\mu_{Ni}$ = -0.70 eV, $\mu_{Fe}$ = -1.09 eV, $\mu_O$ = -2.13 eV. The calculated formation energy of neutral $B_{Ni}^O$ is $\Delta H_f\left(B_{Ni^O}^0\right) = 4.91$ eV.

Figure 5 illustrates the lowest formation energies of substitutional B impurities as a function of $E_F$, calculated for various charge states. For all values of $E_F$ within the band gap, the lowest energy belongs to the substitutional B impurity at the tetrahedral Fe site. If the Fermi level is tuned closer to the VBM, $B_{Fe}^T$ becomes positively charged and its formation energy decreases, reaching the lowest value of approximately 1.3 eV at the VBM ($E_F = 0$).

All of the calculated formation energies of substitutional B impurities in NFO are positive and moderately high. This suggests that the incorporation of B into the NFO structure as a substitutional defect is unlikely, the limiting conditions being the formation of $Ni_3B_2O_6$, $B_2O_3$, or $Fe_3BO_5$ as secondary phases. The compositions of these phases are very close to those considered by Sawicki[9] in a recent Mössbauer analysis of CRUD scrapes from fuel assemblies exhibiting AOA; the compound include iron borate ($FeBO_3$), iron orthoborate ($Fe_3BO_6$) and bonnacordite ($Ni_2FeBO_5$).

*6.5 Interstitial B defects*

As illustrated in Fig. 1, in the spinel structure there are three interstitial sites which can accommodate impurities: two tetrahedral sites, denoted T1 and T2 and one octahedral site, denoted O in Fig. 1. According to Eq. (1), the formation energy of an interstitial B defect can be calculated as $\Delta H_f\left(B_{T/O}^q\right) = \Delta E\left(V_{T/O}^q\right) - \mu_B + qE_F$. To make incorporation of B favorable, $\mu_B$ must be maximized while avoiding formation of secondary phases. Using the calculated formation enthalpies, listed in Table 1, within the limits of NFO stability domain shown in Fig. 3, we find that the maximum possible $\mu_B$ is -3.63 eV and it is limited by the formation of $Fe_3BO_5$. Figure 6 illustrates the formation energies of the interstitial B defects in NFO as a function of $E_F$. The most stable position for B is the tetrahedral interstitial site (T2), with a formation energy that is close to 0.15 eV as the $E_F$ approaches the VBM. Even though this value is considerably lower than all other formation energies, the calculations indicate that it is energetically more favorable for B to form $Fe_3BO_5$ instead of entering the NFO structure as an interstitial impurity.

## 7. Summary

Using first-principles methods, we have investigated the electronic structure of NFO and combined thermodynamical data with theoretical results to analyze the possibility of B incorporation into the spinel structure of NFO. The point defects under investigation include substitutional and interstitial B impurities as well as Ni and Fe vacancies.

Under thermodynamic equilibrium, assuming solid-solid equilibrium between NFO and atomic reservoirs of Ni and Fe, it is unlikely that B is incorporated into the NFO structure. The main factors that limit B incorporation are the narrow chemical potential domain where NFO is stable and the precipitation of various Fe-Ni-B-O compounds as secondary phases. Among these phases the most prevalent appears to be $B_2O_3$, $Fe_3BO_5$, and $Ni_3B_2O_6$.

The incorporation energies depend sensitively on the electron chemical potential ($E_F$) and the charge state of the defect. In n-type NFO, the most stable defect appears to be the Ni vacancy $\left(V_{Ni}^{2-}\right)$ while in p-type material the lowest formation energy belongs to the interstitial B occupying a tetrahedrally coordinated site $\left(B_{T2}^{2+}\right)$. Because of the limiting conditions mentioned above, it is more likely that B will form secondary phases with Fe, Ni and O instead of entering in the NFO structure as a point defect.

### Acknowledgements

This research was supported by the Consortium for Advanced Simulation of Light Water Reactors (http://www.casl.gov), an Energy Innovation Hub (http://www.energy.gov/hubs) for Modeling and Simulation of Nuclear Reactors under U.S. Department of Energy Contract No. DE-AC05-00OR22725. The computational work has been performed at NERSC, supported by the Office of Science of the US Department of Energy under Contract No. DE-AC02-05CH11231.

TABLE 1. Calculated DFT energies per formula unit of the competing and secondary phases used in Eq. (1)-(8) along with the experimental and theoretical enthalpies of formation.

| Compound | DFT energy/f.u. (eV) | $\Delta H^{\mathrm{exp}}$ (eV) | $\Delta H^{\mathrm{theor}}$ (eV) |
|---|---|---|---|
| $B_2O_3$ | -40.13 | -12.83 | -12.85 |
| FeB | -12.01 | -0.76 | -0.38 |
| FeO | -12.83 | -2.73 | -2.45 |
| $Fe_2O_3$ | -34.14 | -8.53 | -8.40 |
| $Fe_3O_4$ | -47.92 | -11.49 | -11.80 |
| NiB | -9.22 | -0.48 | -0.66 |
| $Ni_2B$ | -11.45 | -0.66 | -0.52 |
| $Ni_4B_3$ | -30.01 | -1.86 | -1.95 |
| NiO | -10.16 | -2.48 | -2.83 |
| $Ni_2O_3$ | -24.36 | -5.07 | -4.75 |
| $NiFe_2O_4$ | -44.47 | - | -11.41 |
| $Fe_3BO_5$ | -64.84 | - | -17.55 |
| $Fe_3BO_6$ | -70.77 | - | -18.52 |
| $FeBO_3$ | -37.05 | - | -10.54 |
| $NiB_2O_4$ | -50.34 | - | -15.74 |
| $Ni_3B_2O_6$ | -70.94 | - | -21.69 |
| $Ni_2FeBO_5$ | -57.75 | - | -10.40 |

TABLE 2. The fitted elemental reference energies calculated with Eq. (8). Values are given in eV.

| Element | Fe | Ni | O | B |
|---|---|---|---|---|
| Fitted elemental energy | -5.43 | -2.36 | -4.96 | -6.20 |

TABLE 3. Experimental magnetic moments and theoretical spin magnetic moments (in $\mu_B$) calculated using different Hubbard $U$ parameters.

| | $Fe^T$ | $Fe^O$ | $Ni^O$ | $U_{\mathrm{eff}}(Fe^T)$ | $U_{\mathrm{eff}}(Fe^O)$ | $U_{\mathrm{eff}}(Ni^O)$ |
|---|---|---|---|---|---|---|
| Ref. [45] | -4.86 | 4.73 | 2.22 | - | - | - |
| Ref. [46] | -3.99 | 4.09 | 1.54 | 4.5 | 4.5 | 4.0 |
| Ref. [15] | -3.97 | 4.11 | 1.58 | 3.0 | 3.0 | 3.0 |
| Present work | -4.13 | 4.22 | 1.71 | 4.5 | 4.5 | 6.0 |

TABLE 4. Calculated defect formation energies in NFO for different charge states. In the case of substitutional B at the T2 and O sites no convergent results were obtained for q =-1 and -2. Values are given in eV.

| Defect | Charge state | | | | |
| --- | --- | --- | --- | --- | --- |
|  | q = -2 | q = -1 | q = 0 | q = 1 | q = 2 |
| $V_{Fe}^{T}$ | 9.10 | 8.30 | 7.73 | 7.26 | 6.88 |
| $V_{Fe}^{O}$ | 8.46 | 7.92 | 7.46 | 7.08 | 6.76 |
| $V_{Ni}^{O}$ | 6.13 | 5.56 | 5.12 | 4.75 | 4.41 |
| $B_{Fe}^{T}$ | 3.41 | 1.91 | -0.08 | -0.55 | -0.93 |
| $B_{Fe}^{O}$ | 3.76 | 4.09 | 1.93 | 1.46 | 1.08 |
| $B_{Ni}^{O}$ | 5.54 | 3.46 | 1.94 | -0.08 | -0.53 |
| $B_{T1}$ | 5.03 | 3.17 | 1.45 | -0.32 | -1.08 |
| $B_{T2}$ | - | - | -0.92 | -2.43 | -3.48 |
| $B_{O}$ | - | - | 0.09 | -1.08 | -2.11 |

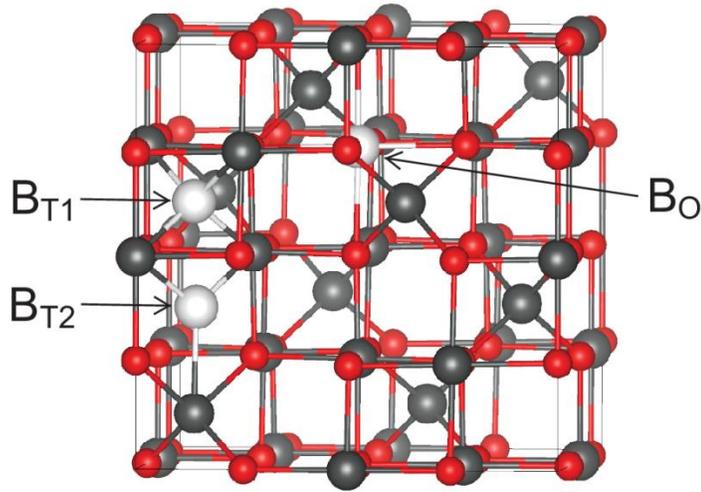

**Figure 1.** Illustration of the spinel structure indicating the two tetrahedral (T1, T2) and octahedral (O) interstitial sites.

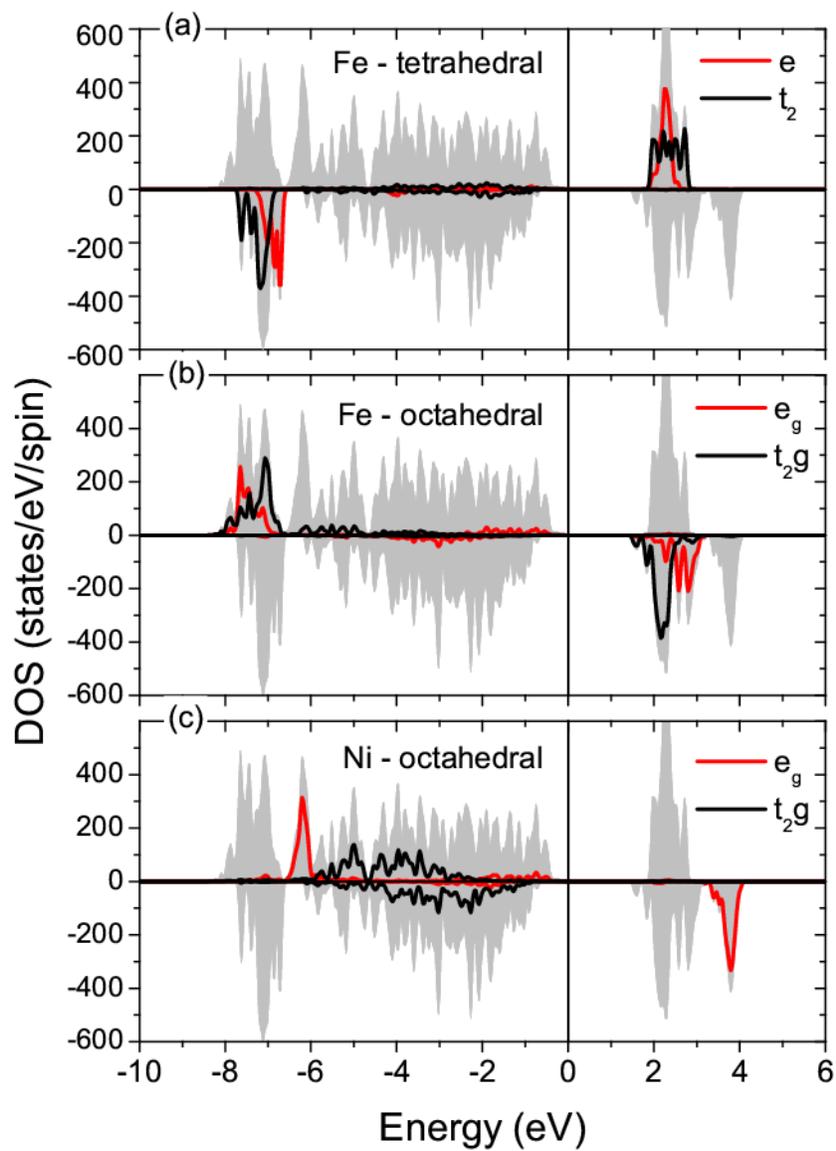

**Figure 2.** Spin polarized DOS of nickel ferrite projected on the Fe and Ni *d* states. Total DOS is also shown as the grey background area. The *d* states of the octahedral cations are separated into $t_{2g}$ and $e_g$ components and the tetrahedral Fe *d* states are separated into $t_2$ and e components.

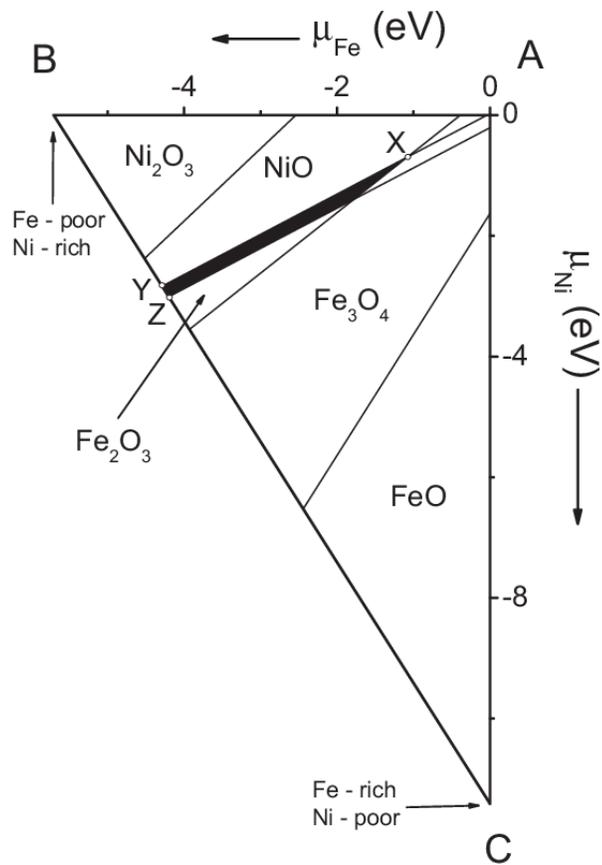

**Figure 3.** Calculated stability domain (dark area) of NFO in the ($\mu_{fe}$, $\mu_{Ni}$) plane. The white regions represent domains of the chemical potentials where secondary phases form.

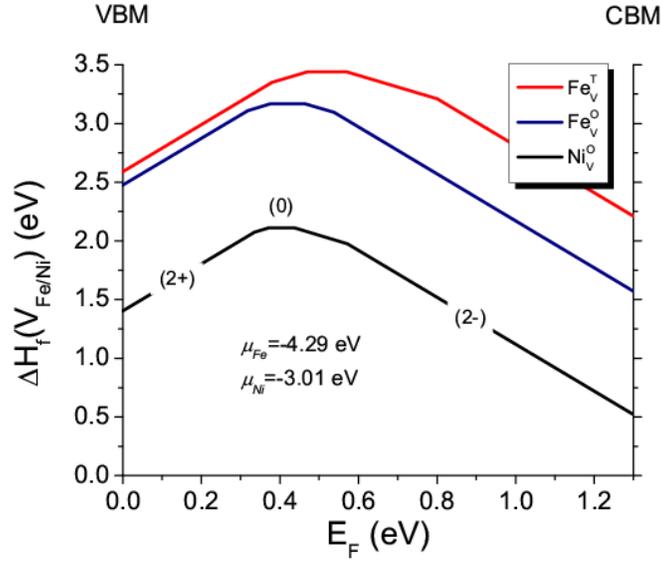

**Figure 4.** Formation energies of Fe and Ni vacancies in NFO as a function of $E_F$, under Fe- and Ni-poor conditions. The slope of the lines represents the charge state of the defect and the value of $E_F$ where the slope changes represents the charge transition level (ionization level). The most stable defect is the Ni vacancy.

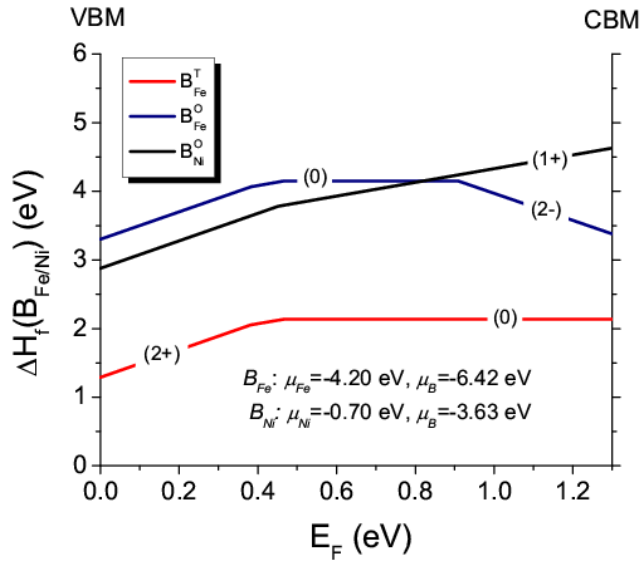

**Figure 5.** Formation energies of substitutional B defects in NFO as a function of $E_F$.

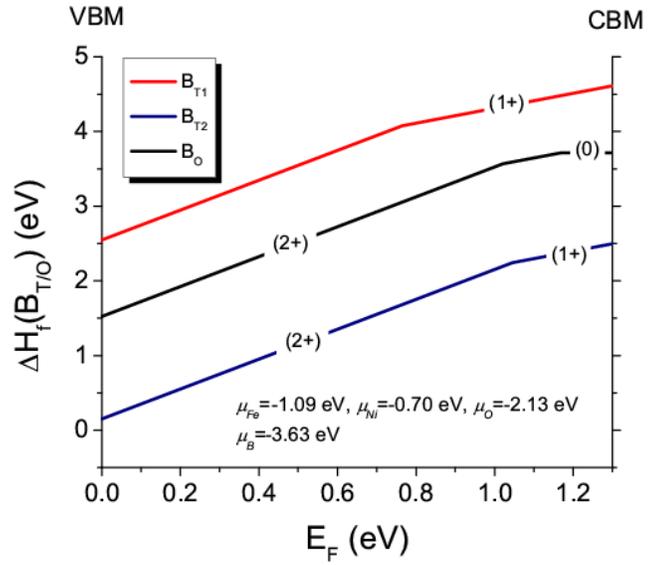

**Figure 6**. Formation energies of substitutional B defects in NFO versus $E_F$. In p-type NFO, the most stable location of B is the tetrahedral interstitial site (T2).